\documentclass[10pt]{iopart}

\usepackage{graphicx}
\usepackage{amssymb}
\usepackage{exscale}
\usepackage{iopams}
\usepackage{subfigure}
\usepackage{xspace}
\usepackage{citesort}

\renewcommand{\Im}[0]{\mathrm{Im}\,}
\renewcommand{\Re}[0]{\mathrm{Re}\,}

\newcommand{\ie}[0]{i.e.\@\xspace}
\newcommand{\eg}[0]{e.g.\@\xspace}

\newcommand{\om}[0]{\omega}

\newcommand{\Ep}{E_\mathrm{p}}
\newcommand{\omb}[0]{\overline{\omega}}

\newcommand{\Ap}[0]{A_\mathrm{p}}

\newcommand{\Ac}[0]{A_\mathrm{c}}
\newcommand{\Aic}[0]{A_\mathrm{ic}}
\newcommand{\nag}{\phantom{\dag}}

\newcommand{\on}{\hat{n}}
\newcommand{\en}{\epsilon}

\newcommand{\las}[0]{\langle}
\newcommand{\ras}[0]{\rangle}

\renewcommand{\tilde}[1]{\widetilde{#1}}

\begin{document}

%%%%%%%%%%%%%%%%%%%%%%%%%%%%%%%%%%%%%%%%%%%%%%%%%%%%%%%%%%%%%%%%%%%%%%%%%%%%%
%%%%%%%%%%%%%%%%%%%%%       TITLE & ABSTRACT          %%%%%%%%%%%%%%%%%%%%%%%
%%%%%%%%%%%%%%%%%%%%%%%%%%%%%%%%%%%%%%%%%%%%%%%%%%%%%%%%%%%%%%%%%%%%%%%%%%%%%
\title{Phonon spectral function of the Holstein polaron}

\author{J Loos,$^1$ M Hohenadler,$^2$ A Alvermann,$^3$ and H Fehske$^3$}

\address{$^1$%
  Institute of Physics, Academy of Sciences of the Czech Republic, Prague
}

\address{$^2$%
  Institute for Theoretical and Computational Physics, TU Graz, Austria
}

\address{$^3$%
  Institute for Physics, Ernst-Moritz-Arndt University Greifswald, Germany
}

\ead{\mailto{loos@fzu.cz}}

\begin{abstract}
  The phonon spectral function of the one-dimensional Holstein model is
  obtained within weak and strong-coupling approximations based on
  analytical self-energy calculations. The characteristic excitations found
  in the limit of small charge-carrier density are related to the known
  (electronic) spectral properties of Holstein polarons such as the polaron
  band dispersion. Particular emphasis is laid on the different physics
  occurring in the adiabatic and anti-adiabatic regimes, respectively.
  Comparison is made with a cluster approach exploiting exact numerical
  results on small systems to yield an approximation for the thermodynamic
  limit. This method, similar to cluster perturbation theory, confirms the
  analytical findings, and yields accurate results also in the
  intermediate-coupling regime.
\end{abstract}


% 71.38.-k : Polarons and electron-phonon interactions
% 63.20.Kr : Phonon-electron and phonon-phonon interactions
% 63.20.Dj : Phonon states and bands, normal modes, and phonon dispersion
% 71.38.Fp : Large or Frohlich polarons
% 71.38.Ht : Self-trapped or small polarons

%
%
%
%%%%%%%%%%%%%%%%%%%%%%%%%%%%%%%%%%%%%%%%%%%%%%%%%%%%%%%%%%%%%%%%%%%%
\section{Introduction}\label{sec:intro}
%%%%%%%%%%%%%%%%%%%%%%%%%%%%%%%%%%%%%%%%%%%%%%%%%%%%%%%%%%%%%%%%%%%%%
%
%
%

Intermediate or strong electron-phonon (EP) interaction gives rise to the
existence of polaronic carriers in a number of interesting materials (see,
\eg, \cite{HaMaDeLoKo04,HaMaLoKo04,BaCeClDeGuKoAeBeFoAD05}). As a
consequence, models for such quasiparticles, \eg, the Holstein model
\cite{Ho59a} considered here, have been investigated intensively in the past
decades in order to understand the process of polaron formation.  Whereas
valuable insight into the latter may be gained by considering a single charge
carrier (\ie, the Holstein polaron problem), real materials are usually
characterized by finite carrier densities, motivating studies of many-polaron
models \cite{Al92,TeDe01,HoNevdLWeLoFe04,DaDaYa05}.

Over the last decade, a large number of theoretical studies, the most
reliable of which based on unbiased numerical methods, have led to a fairly
complete understanding of the Holstein polaron concerning both ground-state
and spectral properties. Calculations of the latter, \eg, the one-electron
Green function, are particularly rewarding as they provide detailed insight
into the non-linear process of an electron becoming self-trapped in the
surrounding lattice distortion. For reviews of the Holstein polaron problem
see \cite{AlMo95,FeAlHoWe06}.

In this paper, we contribute to a completion of the knowledge about the
single-polaron problem by investigating the phonon spectral function, an
important observable rarely considered in previous work.  On the
contrary, it has been studied quite intensively for the spinless
Holstein model and the Holstein-Hubbard model, both at half filling, for
which phonon softening at the zone boundary has been found to occur at the
Peierls transition in one dimension
\cite{SyHuBeWeFe04,SyHuBe05,CrSaCa05,HoWeBiAlFe06,NiZhWuLi05,SyHueBe06}. Note that the
assumption of a local self-energy frequently used in combination with
dynamical mean field theory (DMFT)---as appropriate for infinite
dimensions---leads to an unrealistic wavevector-independent softening of all
phonon excitations \cite{MeHeBu02,KoMeHe04}. The renormalization of the
phonon modes in small clusters with one electron, especially their softening
at the critical EP coupling, has been investigated numerically in
\cite{AlKaRa94}, and results for the coherent phonon spectrum of the infinite
system have been reported in \cite{Ba06}. Furthermore, analytical approximations for the phonon
self-energy and the frequencies of the vibration modes in the coupled EP
system have been obtained in \cite{EnSc63,AlSc97,Al92}.

Here we present analytical calculations valid at weak and strong EP coupling,
respectively, as well as a cluster approach, similar to cluster perturbation
theory \cite{SePePL00,SePePl02,HoAivdL03},
which yields accurate results in all relevant parameter regimes. Both
approaches are capable of describing the momentum dependence of the phonon
spectral functions, and we find a very good agreement between analytical and
numerical results.

Despite the formal restriction to the one-electron case, the numerical
results will correspond to finite electron densities due to the finite
underlying clusters. The analytical calculations presented here are based on
the electron (polaron) spectral functions previously deduced in \cite{LoHoFe06} for the
weak-coupling (WC) and strong-coupling (SC) cases and for charge carrier
concentrations $n<0.5$. These spectral functions depend on $n$ and their
calculation requires a self-consistent determination of the chemical
potential $\mu$. To avoid this difficult task, we shall consider the
analytical formulas in the mathematical limit of $\mu$ approaching the bottom
of the electron (polaron) band.  The resulting limiting shape of the spectral
functions will provide us with a picture of the phonon spectrum for small
carrier concentrations. In particular, as demonstrated below, the positions
of the low-energy excitations will be found to be in a very good agreement
with numerical results. Moreover, the general analytical treatment will
enable us to discuss to some extent the features which occur for
non-negligible carrier concentrations.

The paper is organized as follows. In section~\ref{sec:model}, we introduce
the Holstein model. Section~\ref{sec:methods} is devoted to the derivation of
the analytical results and a discussion of our cluster approach, whereas
section~\ref{sec:results} contains the results and a discussion. Finally, we
summarize in section~\ref{sec:conclusions}.

%
%
%%%%%%%%%%%%%%%%%%%%%%%%%%%%%%%%%%%%%%%%%%%%%%%%%%%%%%%%%%%%%%%%%%%%%
\section{Model}\label{sec:model}
%%%%%%%%%%%%%%%%%%%%%%%%%%%%%%%%%%%%%%%%%%%%%%%%%%%%%%%%%%%%%%%%%%%%%
%
%
%

In view of the SC approximation presented below, it is
convenient to write the Hamiltonian of the one-dimensional (1D) spinless
Holstein model as
\begin{equation}\label{eq:H}
  H = -\mu\sum_j \on_j      
      +\om_0\sum_j b^\dag_j b^{\nag}_j 
      -\sum_{jj'} C^{\nag}_{j'j} c^\dag_{j'}c^{\nag}_j
  \,,
\end{equation}
where
\begin{equation}\label{eq:coeffs}
  C_{jj} = g\om_0 (b^\dag_j + b^{\nag}_j)
  \,,\quad 
  C_{\las j'j \ras} = t
  \,.
\end{equation}
In equation~(\ref{eq:H}), $c^\dag_i$ ($b^\dag_i$) creates a spinless fermion
(a phonon of energy $\om_0$) at site $i$, and $\on_i=c^\dag_i c^{\nag}_i$
with $n_i=0,1$. The first term contains the chemical potential $\mu$ and
determines the carrier density $n$, whereas the second term accounts for the
elastic and kinetic energy of the lattice. Finally, the last term describes
the local coupling between the lattice displacement $x_i=b^\dag_i+b^{\nag}_i$
and the electron density $\on_i$ with the coupling parameter $g$ (for $j=j'$,
cf equation~(\ref{eq:coeffs})), as well as electron hopping processes between
neighbouring lattice sites $\las j'j\ras$ with hopping amplitude $t$ (for
$j=j'\pm1$). We take $t$ as the unit of energy, and set the lattice constant
to unity.

From numerous previous investigations \cite{FeAlHoWe06} of this model in the
one-electron case considered here the following picture emerges. At WC, the
ground state consists of a large polaron, corresponding to a self-trapped
electron with a lattice distortion extending over many lattice sites. As the
EP coupling is increased, a cross-over takes place to a small-polaron state,
in which the lattice distortion is essentially localized at the site of the
electron, leading to a substantial increase of the quasiparticle's effective
mass in the intermediate-coupling (IC) and SC regime. Depending on the value
of the adiabaticity ratio $\alpha=\om_0/t$, the small-polaron cross-over
occurs at a critical value of the EP coupling determined by the more
restrictive of the two conditions $\lambda=\Ep/2t\geq 1$ (relevant for
$\alpha\ll1$) or $g^2=\Ep/\om_0\geq 1$ (for $\alpha\gg1$). Here
$\Ep=g^2\om_0$ is the polaron binding energy in the atomic limit defined by
$t=0$.

%
%
%%%%%%%%%%%%%%%%%%%%%%%%%%%%%%%%%%%%%%%%%%%%%%%%%%%%%%%%%%%%%%%%%%%%%
\section{Methods}\label{sec:methods}
%%%%%%%%%%%%%%%%%%%%%%%%%%%%%%%%%%%%%%%%%%%%%%%%%%%%%%%%%%%%%%%%%%%%%
%
%
%

\subsection{Analytical approach}\label{sec:analytical}

The aim of the present treatment is to deduce and to interpret the essential
features of the numerically calculated phonon spectral functions (see
section~\ref{sec:results}). In the following calculations, we shall deal with
coupled equations of motion of the Matsubara Green functions for the
lattice-oscillator coordinates on the one hand, and for the spinless charge
carriers on the other hand. It will be shown that the general features of the
spectral functions may be understood on the basis of the results obtained for
the WC and SC cases, where an approximate treatment is well justified
\cite{HoNevdLWeLoFe04,LoHoFe06}.

\subsubsection{Weak-coupling approximation}\label{sec:methods:wc}

The Matsubara Green function for phonons, defined as
\begin{equation}
  D(m_1\tau_1;m_2\tau_2) 
  = 
  - \las T_\tau x_{m_1}(\tau_1) x_{m_2}(\tau_2)\ras\,,
\end{equation}
fulfills the equation of motion
\begin{eqnarray}\label{eq:eom}\fl
  \frac{1}{2\om_0}\left(\frac{\partial^2}{\partial\tau_1^2}-\om_0^2\right)
  D(m_1\tau_1;m_2\tau_2) 
  &=&\delta_{m_1m_2}\delta(\tau_1-\tau_2)
  \\\nonumber
  &&-
  g\om_0
  \las T_\tau c^{\nag}_{m_1}(\tau_1)c^\dag_{m_1}(\tau_1)x_{m_2}(\tau_2)\ras
  \,,
\end{eqnarray}
assuming $\las x_{m_2}\ras=0$ in the WC regime. The mixed Green function on
the rhs of equation~(\ref{eq:eom}) will be expressed by means of the
generalized fermionic Green function \cite{KaBa62,BBTy62}
\begin{equation}\label{eq:def:G}
  G(n_1\tau_1;n_2\tau_2;U)
  =
  -\frac{1}{\las S\ras}
  \las T_\tau c^{\nag}_{n_1}(\tau_1)c^\dag_{n_2}(\tau_2) S\ras
  \equiv
  G(1;2;U)
\end{equation}
and
\begin{equation}\label{eq:def:S}
  S = T_\tau
  \exp\left(-\int_0^\beta\rmd\tau\sum_{nn'} U_{nn'}(\tau)
  C_{nn'}(\tau)\right)\,,
\end{equation}
where the classical variables $U_{nn'}(\tau)$ were introduced as a purely
formal device. Consequently, the relation between the mixed Green function
and the fermionic Green function reads
\begin{equation}\fl\quad\quad\quad
  \las T_\tau c^{\nag}_{n_1}(\tau_1)
  c^\dag_{n_1'}(\tau_1')C_{m_2m_2}(\tau_2)\ras
  =
  \left[\frac{\delta}{\delta
      U_{m_2m_2}(\tau_2)}G(n_1\tau_1;n_1'\tau_1';U)\right]_{U=0}
  \,.
\end{equation}
Here $\delta/\delta U$ denotes the functional derivative.
Defining the inverse Green function $G^{-1}(1;2;U)$ to $G(1;2;U)$ according
to \cite{KaBa62} we obtain
\begin{eqnarray}\fl\label{eq:fermionic}
  \las T_\tau c^{\nag}_{m_1}(\tau_1)c^\dag_{m_1}(\tau_1)
  C_{m_2m_2}(\tau_2)\ras
  \\\nonumber
  =
  -\int_0^\beta\rmd\tau'' \sum_{m''} G(m_1\tau_1;m''\tau'')\\\nonumber
  \quad\times
  \int_0^\beta\rmd\tau'  \sum_{m'}
  \left[
     \frac{\delta}{\delta U_{m_2m_2}(\tau_2)}G^{-1}(m''\tau'';m'\tau';U)
  \right]_{U=0}
  G(m'\tau';m_1\tau_1)
\end{eqnarray}
for the interaction term in the equation of motion~(\ref{eq:eom}). The
resulting equation for the phonon Green function is quite general, as no
approximations have been made up to now.

In the sequel, the fermionic Green functions in equation~(\ref{eq:fermionic})
will be obtained using the fermionic spectral functions $A(k,\om)$ which have
been calculated to second order in the EP coupling constant in
\cite{LoHoFe06}. To the same order, the functional derivative of $G^{-1}$
will be determined using the relation between the inverse Green function and
the self-energy $\Sigma$, $G^{-1}(1;2;U)=G^{-1}_0(1;2)-\Sigma(1;2;U)$, where
the Green function of the zeroth order, $G_0$, is independent of $U$.
Accordingly, the derivative of $G^{-1}$ in equation~(\ref{eq:fermionic}) will
be expressed as the derivative of $\Sigma$, which gives to second order
\cite{LoHoFe06,FeLoWe97,FeLoWe00}
\begin{equation}\fl
  \left[
    \frac{\delta}{\delta
      U_{m_2m_2}(\tau_2)}\Sigma(m''\tau'';m'\tau';U)
  \right]_{U=0}
  =
  \las T_\tau
  C_{m''m'}(\tau'')C_{m_2m_2}(\tau_2)\ras\delta_{m''m'}\delta(\tau''-\tau')
  \,.
\end{equation}
Therefore, equation~(\ref{eq:eom}) acquires the following form:
\begin{eqnarray}\label{eq:eom_2}\fl
  \frac{1}{2\om_0}\left(\frac{\partial^2}{\partial\tau_1^2}-\om_0^2\right)
  D(m_1\tau_1;m_2\tau_2)
  &=&
  \delta_{m_1m_2}\delta(\tau_1-\tau_2)
  -
  \int_0^\beta\rmd\tau' \sum_{m'} G(m_1\tau_1;m'\tau')\\\nonumber
  &&\times
  \las T_\tau C_{m'm'}(\tau') C_{m_2m_2}(\tau_2)\ras G(m'\tau';m_1\tau_1)
  \,.
\end{eqnarray}
However, 
\begin{equation}\label{eq:ferm_phon}\fl\quad
  \las T_\tau C_{m'm'}(\tau')C_{m_2m_2}(\tau_2)\ras 
  =
  g^2\om_0^2\las T_\tau x_{m'}(\tau') x_{m_2}(\tau_2)\ras
  =
  -g^2\om_0^2 D(m'\tau';m_2\tau_2)
  \,.
\end{equation}
Substitution of equation~(\ref{eq:ferm_phon}) into equation~(\ref{eq:eom_2})
and subsequent multiplication on the rhs by $D^{-1}$ give
\begin{equation}\label{eq:pi}
  \Pi(m_1\tau_1;m_2\tau_2) 
  = 
  g^2\om_0^2 G(m_1\tau_1;m_2\tau_2) G(m_2\tau_2;m_1\tau_1)
\end{equation}
with $\Pi = D_0^{-1} - D^{-1}$, the self-energy of the
phonons in the Matsubara framework.

Fourier transformation of equation~(\ref{eq:pi}) leads to
\begin{equation}\label{eq:pi2}
  \Pi(q,\rmi \om_n)
  =
  g^2\om_0^2\frac{1}{N}\sum_k\frac{1}{\beta}\sum_\nu
  G(k,\rmi\om_\nu)
  G(k+q,\rmi(\om_\nu+\om_n))\,,
\end{equation}
where $\om_\nu = (2\nu+1)\pi/\beta$ and $\om_n = 2n\pi/\beta$
are Matsubara frequencies for fermions and bosons, respectively.

The substitution of the spectral representation of the fermionic Green
functions 
\begin{equation}\label{eq:spectral-theorem}
  G(k',\rmi\om_\nu) =
  \int_{-\infty}^\infty\,\rmd\om'\frac{A(k',\om')}{\rmi\om_\nu-\om'}
\end{equation}
into equation~(\ref{eq:pi2}), the summation over the frequencies $\om_\nu$,
the analytical continuation $\rmi\om_n\mapsto\omb=\om+\rmi\delta$ and the
subsequent limit $\delta\to0$ lead to
\begin{eqnarray}\nonumber
    \Pi(q,\om+\rmi0^+)
    &=&
    g^2\om_0^2\frac{1}{N}\sum_k\int_{-\infty}^{\infty}\rmd\om'
    \int_{-\infty}^\infty \rmd\en' A(k,\om')\\\label{eq:pi3}
    &&\times
    A(k+q,\en')
    [f(\om')-f(\en')]\zeta(\om+\om'-\en')
\end{eqnarray}
with
\begin{equation}
  f(x) = \frac{1}{\rme^{\beta x} + 1}\,,\quad 
  \zeta(x)
  =
  \frac{\mathcal{P}}{x}-\rmi\pi\delta(x)
  \,.
\end{equation}
In the low-temperature limit the integrand of equation~(\ref{eq:pi3}) may be
non-zero only if 
\begin{eqnarray}\nonumber
\om'<0\; &\mathrm{and}\;& \en'>0 \Leftrightarrow f(\om')-f(\en')=1\quad\mathrm{or}\\
\om'>0\; &\mathrm{and}\;& \en'<0 \Leftrightarrow f(\om')-f(\en')=-1
\,.
\end{eqnarray}
Using the second-order result for the electron spectral function $A(k',\om')$
deduced in \cite{LoHoFe06}, the calculation of the phonon
self-energy~(\ref{eq:pi3}), the corresponding retarded phonon Green function
$D^\mathrm{R}(q,\om)$ and the phonon spectral function $B(q,\om)$ would be
straight forward \cite{Ma90}. However, $A(k',\om')$ as derived in
\cite{LoHoFe06} depends on the charge-carrier concentration $n$ and has to be
determined self-consistently with a condition for the chemical potential
$\mu$. The situation is thus simplified if we restrict ourselves to the case
of small carrier concentration, for which the dependence on $n$ is expected
to be unimportant. To choose the dominant contributions to the integral on
the rhs of equation~(\ref{eq:pi3}) for small $n$, the integration over $\om'$
and $\en'$ will be divided according to the character of the electronic
spectral functions $A(k,\om')$ and $A(k+q,\en')$.

The coherent part of the spectrum $\Ac(k,\om)$, non-zero for $|\om|<\om_0$,
consists of quasiparticle peaks,
\begin{equation}\label{eq:coh}
\Ac(k,\om) = z_k \delta(\om-(E_k-\mu))\,,
\end{equation}
whereas outside this frequency interval the incoherent spectral function
$\Aic(k,\om)$ is formed by peaks of finite width. In the sequel, the
integrals obtained in this way will be examined with respect to the behaviour
in the limit of small concentrations, \ie, for $\mu$ lying near $E_0$, the
bottom of the band defined by equation~(\ref{eq:coh}).

The real and imaginary parts of $\Pi(q,\om)$ are obtained if the real and
imaginary parts of the $\zeta$-function are substituted into
equation~(\ref{eq:pi3}). The behaviour of the resulting integrals for $\mu$
in the neighbourhood of $E_0$ is then analyzed in the mathematical limit
$\mu\to E_0^+$. We find that the integration for $\Re\Pi(q,\om)$ according to
equation~(\ref{eq:pi3}) gives zero in the limit of vanishing carrier
concentration. In contrast, the integration for $\Im\Pi(q,\om)$ yields a
non-zero result in this limit, with the coherent parts of the spectra,
$\Ac(k,\om')$ and $\Ac(k+q,\en')$, being the only non-vanishing
contributions. Taking into account only the latter, we shall express
$\Im\Pi(q,\om)$ as
\begin{eqnarray}\label{eq:im_Pi}\nonumber
  \Im\Pi(q,\om)
  &=&
  -\frac{1}{2}g^2\om_0^2\int_{-\pi}^\pi\rmd k z_k z_{k+q}
  \delta(\om-(E_{k+q}-E_k)) \theta(\mu-E_k)\\\nonumber
  &&\times
 \int_0^{\om_0} \rmd\en' 
  \delta(\en'-(E_{k+q}-\mu))\\\nonumber
  &=&
  -\frac{1}{2}g^2\om_0^2\int_{|k|<k_\mathrm{F}} \rmd k z_k z_{k+q} 
  \theta(\mu-E_k)
  \frac{\delta(k-k_0)}{|E'_{k_0}-E'_{k_0+q}|}
  \\
  &&\times
  \int_0^{\om_0} \rmd\en'
  \delta(\en'-(E_{k+q}-\mu))
  \,,
\end{eqnarray}
with $E_k\leq\mu<E_{k+q}<\om_0$.
According to \cite{LoHoFe06}, the energies of the electronic band are given
by the equation
\begin{equation}\label{eq:band_WC}
  E_k = \xi_k
  + 
  \frac{g^2\om_0^2}{\pi W}\mathcal{P}\int_{-W}^W\frac{\rmd\xi}{\sqrt{1-(\xi/W)^2}}
  \frac{1}{E_k-\om_0-\xi}
  \,,
\end{equation}
where $\xi_k=-W\cos k$ ($W=2t$), and its derivative
\begin{equation}
  E_k'=W z_k \sin k\,,
\end{equation}
with the spectral weight \cite{LoHoFe06}
\begin{equation}\label{eq:wc_z_k}
  z_k 
  =
  \left|
    1 + \frac{g^2\om_0^2}{\pi W}\mathcal{P}\int_{-W}^W
    \frac{\rmd\xi}{\sqrt{1-(\xi/W)^2}}
    \frac{1}{(E_k-\om_0-\xi)^2}
  \right|^{-1}
  \,.
\end{equation}
The wavevector $k_0$ in equation~(\ref{eq:im_Pi}) lies in the neighbourhood of
$k=0$ defined by $\theta(\mu-E_k)=1$ and satisfies
\begin{equation}\label{eq:om}
  \om-E_{k+q}+E_k = 0
  \,.
\end{equation}
Consequently, the value of $k_0$ appears to be a function of $\om$ at fixed
wavevector $q$, fulfilling $E_{k_0}<\mu$.

To obtain a qualitative picture of the phonon spectral function, we
explicitly take the limit $\mu\to E^+_0$. Thereby, $\lim_{\mu\to
  E^+_0}\Re\Pi(q,\om)= 0$ and the rhs of equation~(\ref{eq:im_Pi}) is
non-zero only on the curve $\om=\om_q=E_q-E_0$ provided that $0<\om<\om_0$.
The values of $\Im \Pi(q,\om)$ are given as follows:
\begin{equation}\label{eq:im_pi}\fl\quad\quad\quad
  \Im\Pi(q,\om) = 
  -\frac{1}{2}g^2\om_0^2\frac{z_0}{W|\sin q|}\theta(\om_0-(E_q-E_0))
  \Delta(\om-(E_q-E_0))
  \,,
\end{equation}
where the function $\Delta(x_1-x_2)$ is the generalization of the Kronecker
symbol being equal to unity for $x_1=x_2$ and zero otherwise. Consequently,
in this limit, the narrow peaks of finite frequency width following for
$\Im\Pi(q,\om)$ from equation~(\ref{eq:im_Pi}), are replaced by discrete
lines given by equation~(\ref{eq:im_pi}). The divergence of the imaginary
part of the phonon self-energy at $q=0$ and $q=\pi$ is connected with the 1D
electron band dispersion. Our method of calculation only takes into account
contributions to the electronic spectral function up to second order, which
is insufficient if divergences occur.

On physical grounds, one expects no
renormalization of the phonon excitations in the zero-density limit, \ie, for
one electron in an infinite lattice. The non-zero imaginary part of the
phonon self-energy even for $n\to 0$ [equation~(\ref{eq:im_pi})] does not
contradict this expectation since the integrated weight of the corresponding
features is zero. As discussed in section~\ref{sec:results}, these non-zero
contributions to the limit of the phonon spectrum may be related to results
for small but finite carrier densities.

So far, we have restricted ourselves to $\om>0$, but the case $\om<0$ may be
treated quite analogously. The only non-zero contribution to $\Im\Pi(q,\om)$
in the limit $\mu\rightarrow E_0^+$ is obtained for $E_{k+q}<\mu<E_k<\om_0$
in the frequency range $-\om_0<\om<0$ on the curve $\om_q=-(E_q-E_0)$, with
the result
\begin{equation}
  \Im\Pi(q,\om) = 
  -\Im\Pi(q,|\om|)
  \,.
\end{equation}

The retarded Green function $D^\mathrm{R}(q,\om+\rmi\delta)$ as the
analytical continuation of
\begin{equation}\label{eq:dqw}
  D(q,\rmi\om_n)
  =
  \frac{2\om_0}
  {(\rmi\om_n)^2 - \om_0^2-2\om_0\Pi(q,\rmi\om_n)}
\end{equation}
in the upper complex half-plane determines the phonon spectral function
\begin{equation}
  B(q,\om)
  =
  -\frac{1}{\pi}
  \Im D^\mathrm{R}(q,\om+\rmi0^+)
  \,.
\end{equation}
Using equation~(\ref{eq:dqw}) and the preceding analysis of $\Pi(q,\om+\rmi0^+)$
in the limit $\mu\rightarrow E_0^+$, we may conclude that for $\om>0$
\begin{equation}\label{eq:B<}
  B(q,\om) 
  =
  -\frac{1}{\pi}\frac{(2\om_0)^2\Im\Pi(q,\om)}{(\om^2-\om_0^2)^2+[2\om_0\Im\Pi(q,\om)]^2}
\end{equation}
if $\om=E_q-E_0$ and $0<\om<\om_0$, where $\Im\Pi$ is given by
equation~(\ref{eq:im_pi}). Otherwise $\Im\Pi(q,\om>0)=0$ and
\begin{equation}\label{eq:B>}
  B(q,\om) = \delta(\om-\om_0)
  \,.
\end{equation}
The result for $\om<0$, analogous to equations~(\ref{eq:B<}),~(\ref{eq:B>})
fulfills the relation
\begin{equation}\label{eq:Bqw_symmetry}
  B(q,\om) = - B(q,-\om)
  \,,
\end{equation}
in agreement with the general requirement on the imaginary parts of retarded
Green functions of real dynamical variables \cite{Zubarev74}.

\subsubsection{Strong-coupling approximation}\label{sec:methods:sc}

The equation of motion~(\ref{eq:eom}) does not hold exactly in the SC regime,
as the coordinate $x_m$ of the local oscillator at site $m$ implies a shift
due to the local lattice deformation associated with on-site small-polaron
formation. In fact, $\las x_{m}\ras=2g\las c^\dag_{m} c^{\nag}_{m}\ras$,
which follows from the Lang-Firsov canonical displacement transformation
\cite{LangFirsov}. However, dealing with the limit of negligible
charge-carrier concentration, $\las x_{m}\ras=0$ may again be assumed. On the
other hand, the charge-carrier number operator in the electron picture is
equal to the number operator in the small-polaron picture. Therefore, we
interpret the Fermi operators $c^{\nag}_{m_1}(\tau_1)$,
$c^\dag_{m_1}(\tau_1)$ in equation~(\ref{eq:eom}) as annihilation and
creation operators of small polarons---the correct quasiparticles in the SC
limit. Accordingly, the mixed term on the rhs of equation~(\ref{eq:eom}) will
be expressed using the generalized small-polaron Green functions, defined
again by equations~(\ref{eq:def:G}),~(\ref{eq:def:S}), where the operators
$C_{nn}(\tau)=g\om_0 x_n(\tau)$ as before, but the $C_{\las nn'\ras}$
correspond to the nearest-neighbour hopping term in the SC regime, \ie,
\begin{equation}\label{eq:sc:C}
  C_{\las nn'\ras}=t\exp\{-g(b^\dag_n-b^{\nag}_n-b^\dag_{n'}+b^{\nag}_{n'})\}
  \,.
\end{equation}
The formalism of generalized Green functions of small polarons was introduced
by Schnakenberg \cite{Sc66} and applied to self-energy calculations in
\cite{FeLoWe97,FeLoWe00,Lo94}. Apart from the $C_{\las nn'\ras}$ given by
equation~(\ref{eq:sc:C}), in contrast to previous work we also include
$C_{nn}=g\om_0 x_n$ in our definition of the generalized Green function.

The presence of the coefficients $C_{nn}$ in the time-ordered exponential in
equation~(\ref{eq:def:S}) causes the polaronic operators not to commute with
the exponent due to the oscillator shift proportional to $c^\dag_n
c^{\nag}_n$. Accordingly, the zeroth-order generalized small-polaron Green
function $G_0$, corresponding to the atomic limit $t=0$, is $U$-dependent
because it fulfills the equation of motion
\begin{equation}\fl
  \left(-\frac{\partial}{\partial\tau_1}-\eta-2g^2\om_0
    U_{n_1n_1}(\tau_1)\right)
  G_0(n_1\tau_1;n_2\tau_2;U)
  =
  \delta_{n_1n_2}\delta(\tau_1-\tau_2)
  \,,
\end{equation}
where $\eta=-\mu-g^2\om_0$. Using matrix notation \cite{KaBa62},
\begin{equation}
  G^{-1}_0(1;1';U) = \left(-\frac{\partial}{\partial\tau_1}-\eta-2g^2\om_0
    U_{n_1n_1}(\tau_1)\right)\delta_{n_1n'_1}\delta(\tau_1-\tau'_1)
\end{equation}
represents the matrix inverse to $G_0$.

To obtain a qualitative picture of the phonon spectral function at SC we
consider the limit $g^2\gg 1$. According to previous considerations
\cite{HoNevdLWeLoFe04,LoHoFe06}, the polaron spectral function in this limit
is dominated by the coherent part representing the polaron band of width
$2W\rme^{-g^2}$, showing that small polarons are the correct quasiparticles
and that multi-phonon processes in the self-energy are negligible.
Consequently, the Green functions in equation~(\ref{eq:fermionic}) will be
expressed by means of equation~(\ref{eq:spectral-theorem}) using the coherent
polaron spectral function
\begin{equation}\label{eq:pol_A}
  \Ap(k,\om) = \delta(\om-(\xi_k+\eta))\,,
\end{equation}
where $\xi_k=-\tilde{W}\cos k$ ($\tilde{W}=2t\rme^{-g^2}$) and
\begin{equation}\fl\label{eq:G-G0_SC}
  \left[
    \frac{\delta}{\delta U_{m_2m_2}(\tau_2)} G^{-1}(1;1';U)
  \right]_{U=0}
  =
  \left[
    \frac{\delta}{\delta U_{m_2m_2}(\tau_2)} G_0^{-1}(1;1';U)
  \right]_{U=0}
\end{equation}
is assumed to be a good approximation.

Using equation~(\ref{eq:G-G0_SC}) and substituting equation~(\ref{eq:fermionic})
into equation~(\ref{eq:eom}), 
\begin{eqnarray}\label{eq:eom_sc}\fl
  \frac{1}{2\om_0}\left(\frac{\partial^2}{\partial\tau_1^2}-\om_0^2\right)
  D(m_1\tau_1;m_2\tau_2)
  &=&
  \delta_{m_1m_2}\delta(\tau_1-\tau_2)\\\nonumber
  &&-
  2g^2\om_0G(m_1\tau_1;m_2\tau_2)G(m_2\tau_2;m_1\tau_1)
\end{eqnarray}
is obtained. Fourier transformation of equation~(\ref{eq:eom_sc}), use of the
spectral representation of the polaron Green function based on
equations~(\ref{eq:spectral-theorem}),~(\ref{eq:pol_A}), and summation over
the fermionic Matsubara frequencies result in
\begin{eqnarray}\label{eq:D_SC}\fl
  \quad\quad\quad\quad\quad\quad\quad
  D(q,\rmi\om_n)
  &=&
  \frac{2\om_0}{(\rmi\om_n)^2-\om_0^2}-2g^2\om_0\frac{2\om_0}{(\rmi\om_n)^2-\om_0^2}
  I(q,\rmi\om_n)\\\nonumber
  I(q,\rmi\om_n)
  &=&
  \frac{1}{N}\sum_k\int_{-\infty}^\infty\,\rmd\om'
  \int_{-\infty}^\infty\,\rmd\en'
  A(k,\om')A(k+q,\en')
  \frac{f(\om')-f(\en')}{\om'-\en'+\rmi\om_n}
  \,.
\end{eqnarray}
The analytical continuation $\rmi\om_n\mapsto\omb=\om+\rmi\delta$ into the
upper complex half-plane gives the retarded Green function
$D^\mathrm{R}(q,\omb)$ and, according to the argument at the end of the
preceding section, only $\om>0$ is to be considered.  After the analytical
continuation, the integral $I(q,\om+\rmi0^+)$ on the rhs of
equation~(\ref{eq:D_SC}) is analogous to the integral in
equation~(\ref{eq:pi3}), but the coherent spectrum (\ref{eq:pol_A}) is
limited to a frequency interval well below the value $\om=\om_0$. The
equations determining $\Im I(q,\om)$ are simplified compared to
equations~(\ref{eq:im_Pi})\,--\,(\ref{eq:om}) since, according to the
small-polaron spectral function~(\ref{eq:pol_A}) used, the spectral weight
$z_k=1$ and the band energy $E_k=\xi_k-g^2\om_0$. The limiting procedure
described in section~\ref{sec:methods:wc} gives again $\Re I(q,\om)=0$ and
\begin{equation}
  \Im I(q,\om) = -\frac{\Delta(\om-(\xi_q-\xi_0))}{2\tilde{W}|\sin
    q|}
  \,.
\end{equation}
Consequently, the mathematical limit of $\Im D^\mathrm{R}(q,\om)$ for
$\mu\to\xi(0)-g^2\om_0$ yields the limiting shape of the SC phonon spectral
function as
\begin{equation}\label{eq:Bqw_SC}
  B(q,\om) = \delta(\om-\om_0)
  + \frac{2g^2\om^2_0}{\om_0^2-\om^2}
  \frac{\Delta(\om-\tilde{W}(1-\cos q))}{\pi\tilde{W}|\sin q|}
  \,.
\end{equation}
The second term on the rhs of equation~(\ref{eq:Bqw_SC}) reflects the
energies of the small-polaron band which in the SC case lies entirely below
the phonon resonance frequency $\om_0$. The divergences occurring at $q=0$,
$\pi$ are again related to the dispersion of the 1D band and result from the
failure of the approximation used at these wavevectors.

\subsection{Numerical cluster approach}\label{sec:methods:cpt}

The numerical approach used here is similar to cluster perturbation theory
\cite{SePePL00,SePePl02,HoAivdL03} for the one-electron Green function. For
the case of the phonon Green function, it has first been proposed in
\cite{FeAlHoWe06} and applied to the half-filled, spinless Holstein model in
\cite{HoWeBiAlFe06}. A previous numerical cluster study of the
renormalization of phonon excitations can be found in \cite{AlKaRa94}.

The phonon spectral function $B(q,\om)$ is defined by means of the retarded
phonon Green function which determines the response of the lattice to the
external perturbation linearly coupled to the phonon variables
\cite{Zubarev74}. The values of $B(q,\om)$ may be shown to be proportional to
the transition probabilities per unit time (at $T\neq0$ averaged with respect
to the canonical distribution) for the transitions induced by the
perturbation having frequency $\om$.

In our case
\begin{equation}\label{eq:bqw_cpt}
  B(q,\om)= -\frac{1}{\pi}\Im D^\mathrm{R}(q,\om)
\end{equation} 
will be calculated at zero temperature, so that the equality
\begin{equation}\label{eq:phononSF}
  D^\mathrm{R}(q,\om) 
  = 
  \lim_{\eta\to 0^+}\ 
  \las \psi_0 | x_{q} 
  \frac{1}{\om + \rmi\eta - H}
  x_{-q} |\psi_0\ras
\end{equation}
holds for $\om>0$. Here $|\psi_0\ras$ denotes the ground state of the infinite
system, and the phonon coordinates are given by
$x_{q}=\frac{1}{\sqrt{N}}\sum_j x_j e^{-\rmi qj}$. For 
the Holstein model~(\ref{eq:H}) we have $B(\pm q,\om)=B(q,\om)$.

The spectral function~(\ref{eq:bqw_cpt}) fulfils a sum rule of the form
$\int_0^\infty\rmd\om B(q,\om)=\las\psi_0|x_{q}x_{-q}|\psi_0\ras$,
which relates the integrated spectral weight to the lattice elongation
in the ground state \cite{Ma90}. The numerical techniques employed in 
the next sections guarantee that this sum rule holds up to 
machine precision.  Note that the analytical results of
section~\ref{sec:analytical} fulfil the usual sum rule for the phonon
spectral function \cite{Ma90} for $T\to0$.

To proceed, as a first step, we divide the infinite lattice into identical
clusters of $N_\mathrm{c}$ sites each, and calculate the cluster Green
function $D^\mathrm{R,(c)}_{ij}(\omega)$ of the Hamiltonian~(\ref{eq:H}) with
one electron and $\mu=0$ for all non-equivalent pairs of sites
$i,j=1,\dots,N_\mathrm{c}$.  For this purpose, we employ the kernel
polynomial method (KPM). Details about the computation of the Green function
by the KPM and its advantages over the widely used Lanczos method can be
found in \cite{WeWeAlScFe05}. The phonon Hilbert space is truncated
\cite{WeWeAlScFe05} such that the resulting error of the spectra is
negligible ($<10^{-4}$), and we have used 1024 moments for the spectra shown
below.

In cluster perturbation theory, an approximation for $G^\mathrm{R}(k,\om)$ of
the infinite system is obtained by taking into account the first-order
inter-cluster hopping processes, leading to a simple Dyson equation
\cite{SePePl02}. However, in the case of the phonon Green function, it turns
out that the first-order term vanishes, since the electron number per cluster
is conserved \cite{HoWeBiAlFe06}. As a result, the cluster approach used here
reduces to a Fourier transformation of the cluster Green function,
\begin{equation}\label{eq:fourier}
  D^\mathrm{R}(q,\omega)
  =
  \frac{1}{N_\mathrm{c}}
  \sum_{i,j=1}^{N_\mathrm{c}} D^{R,(\mathrm{c})}_{ij}(\omega)
  \rme^{-\rmi q(i-j)}
  \,.
\end{equation}
Nevertheless, it represents a systematic approximation to the exact Green
function, as results improve with increasing cluster size $N_\mathrm{c}$.
Moreover, the method becomes exact both for a non-interacting system ($g=0$),
and in the atomic-- or SC limit $t=0$. We shall see below that, provided
$N_\mathrm{c}$ is large enough to capture the physically relevant non-local
correlations, the method yields accurate results for all interesting
parameters. Note that the defects mentioned in \cite{HoWeBiAlFe06}
originating from the neglect of true long-range order at half filling (above
the Peierls transition) are absent in the low-density limit considered here.

%
%
%%%%%%%%%%%%%%%%%%%%%%%%%%%%%%%%%%%%%%%%%%%%%%%%%%%%%%%%%%%%%%%%%%%%%
\section{Results}\label{sec:results}
%%%%%%%%%%%%%%%%%%%%%%%%%%%%%%%%%%%%%%%%%%%%%%%%%%%%%%%%%%%%%%%%%%%%%
%
%
%

In contrast to previous considerations
\cite{Al92,AlKaRa94,EnSc63,AlSc97,Ba06} focussing on the renormalization of
the vibration modes and their softening in the IC regime, the aim of this
work is the calculation of the $q$-dependent phonon spectral functions in the
entire frequency range and all relevant EP-coupling regimes. Particular
attention is paid to the connection of the low-energy features with the
electron (polaron) spectra studied previously \cite{HoAivdL03,LoHoFe06}. In
the analytical calculations of section~\ref{sec:analytical}, the primary goal
was the determination of the imaginary part of the phonon self-energy
together with an analysis of the coherent and incoherent parts of the
electron (polaron) spectral functions.

\begin{figure*}\centering
  \includegraphics[width=0.52\textwidth]{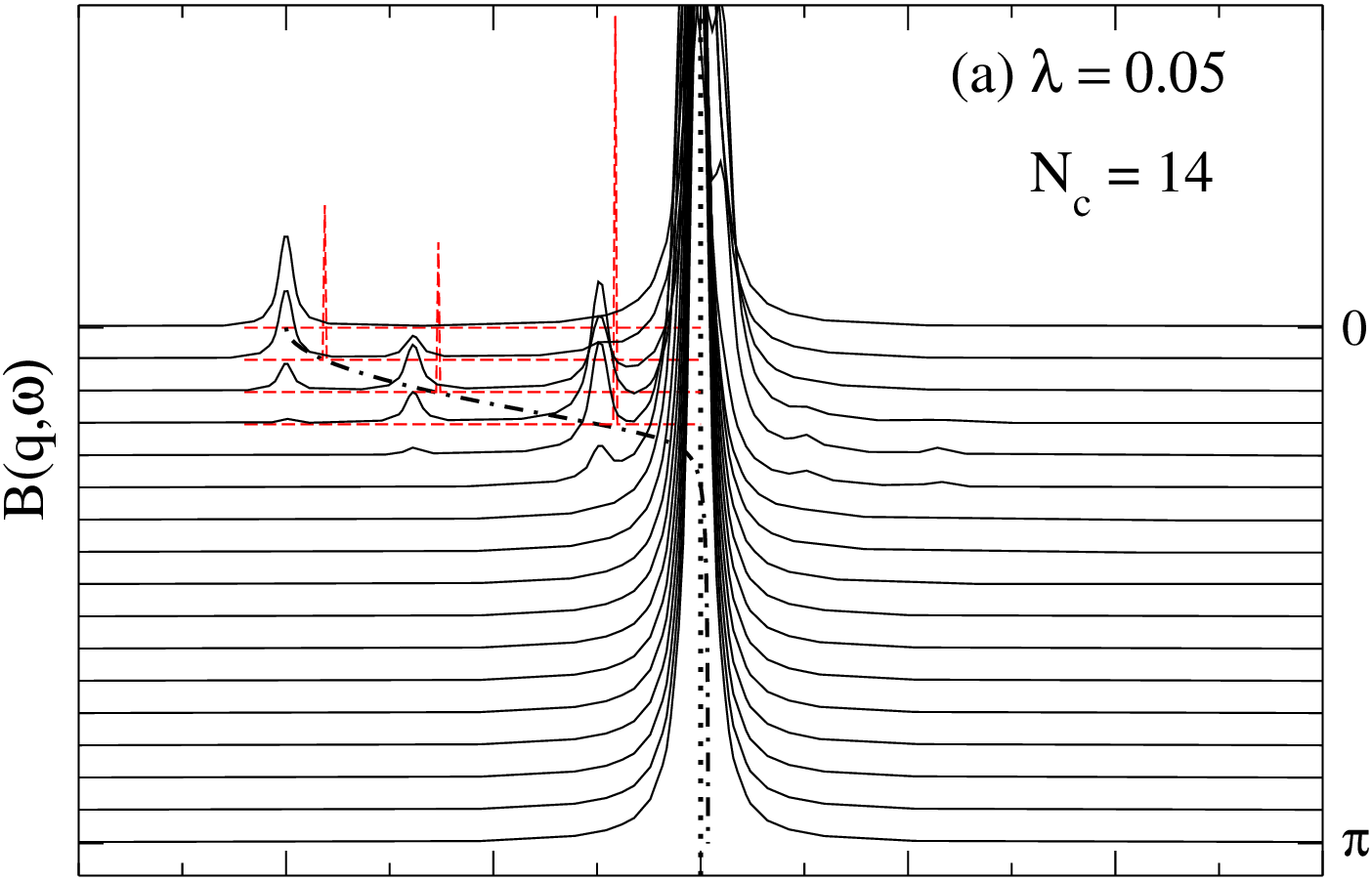}\\
  \includegraphics[width=0.52\textwidth]{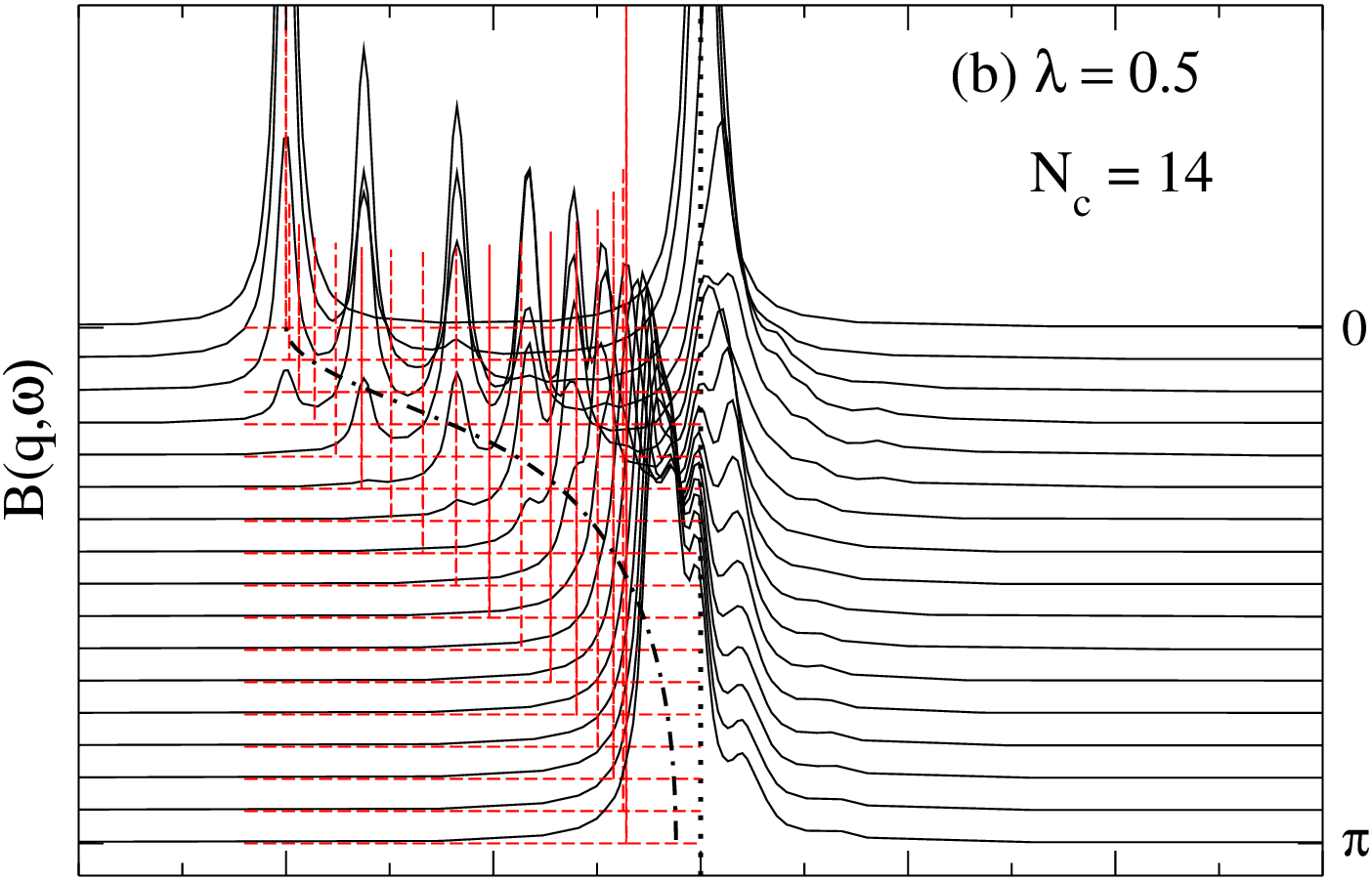}\\
  \includegraphics[width=0.52\textwidth]{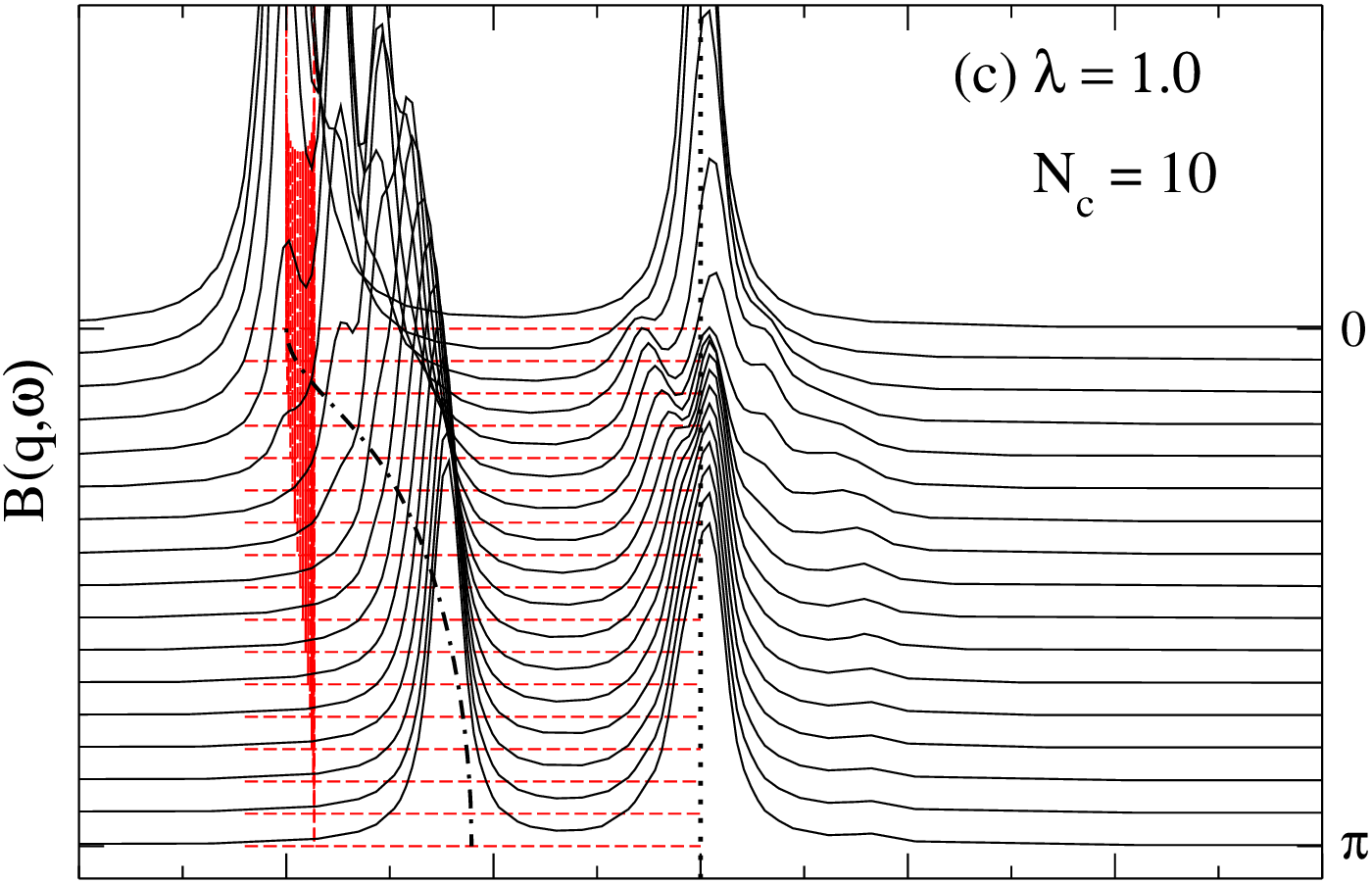}\\
  \includegraphics[width=0.52\textwidth]{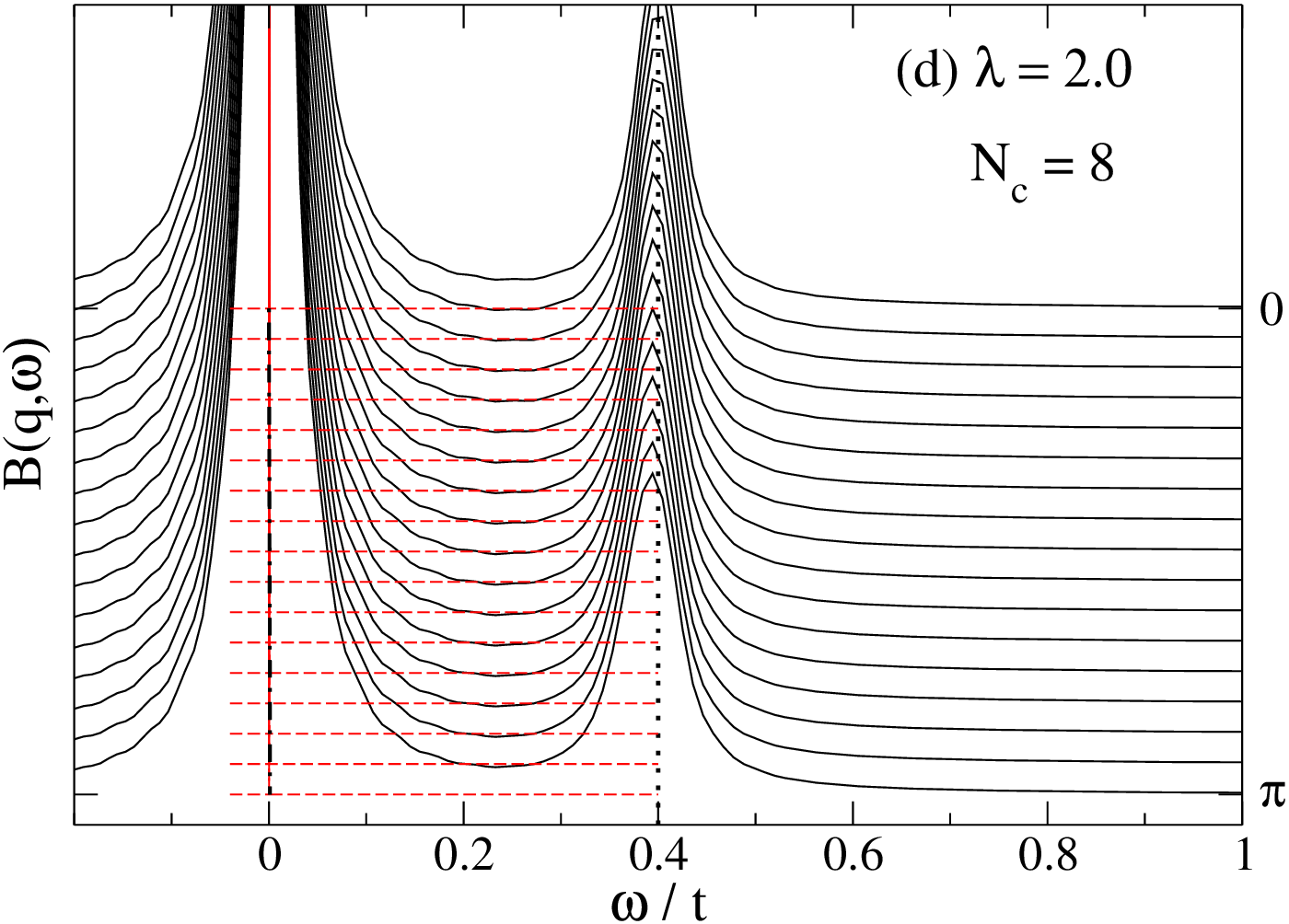}
  \caption{\label{fig:polaron_w0.4} (colour online) Phonon spectral function
    $B(q,\om)$ as obtained by the
    cluster approach, for $\alpha=0.4$ and different EP couplings $\lambda$
    and cluster sizes $N_\mathrm{c}$. WC (a) and SC (b)\,--\,(d) analytical
    results ($\dashed$, see text) have been multiplied by a factor of 10.
    Also shown are the bare phonon frequency $\omega_0$ ($\dotted$), and the
    polaron band dispersion $E_q-E_0$ ($\chain$) as calculated with the
    method of \cite{BoTrBa99}.}
\end{figure*}

\begin{figure*}\centering
  \includegraphics[width=0.52\textwidth]{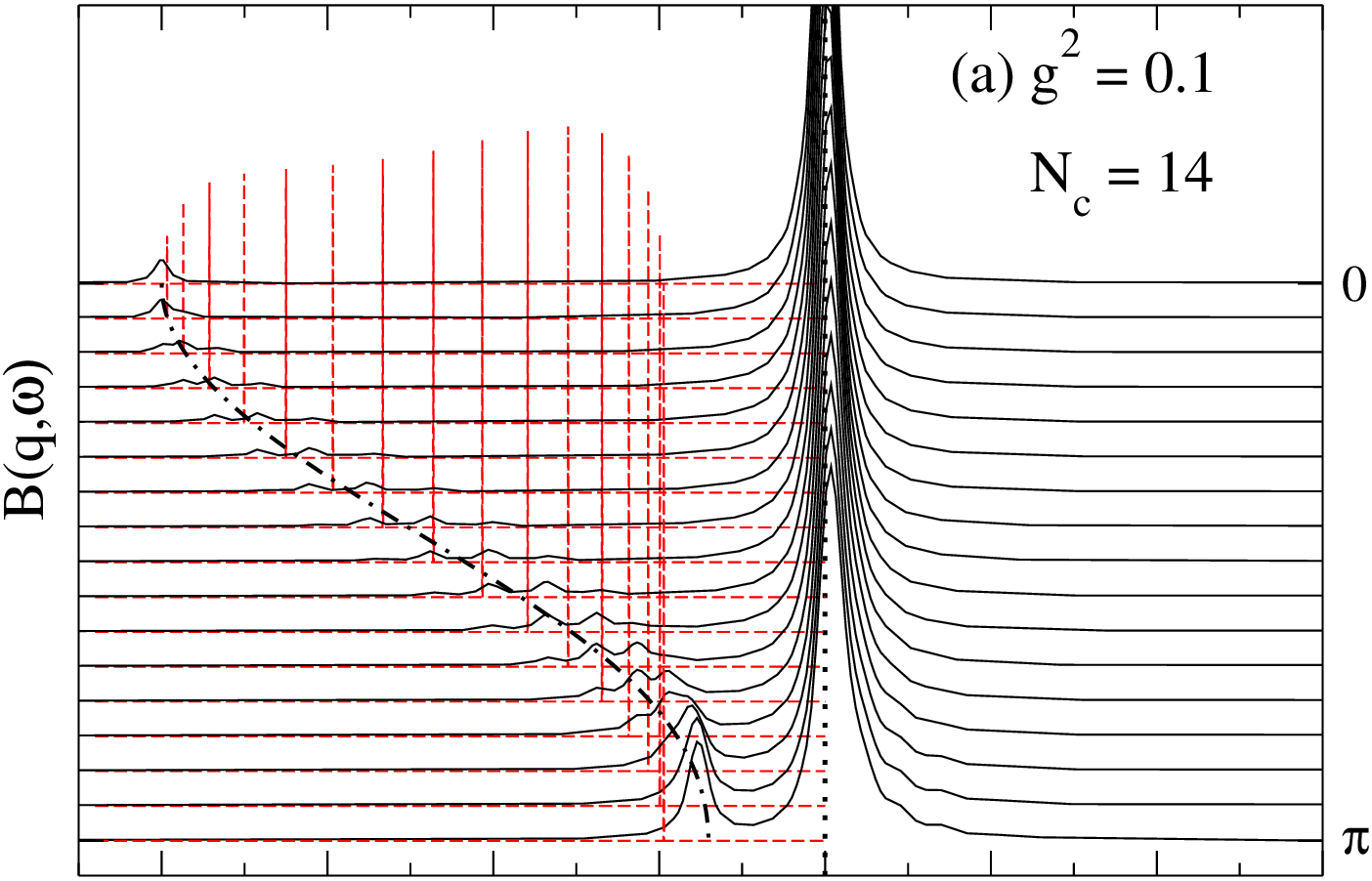}\\
  \includegraphics[width=0.52\textwidth]{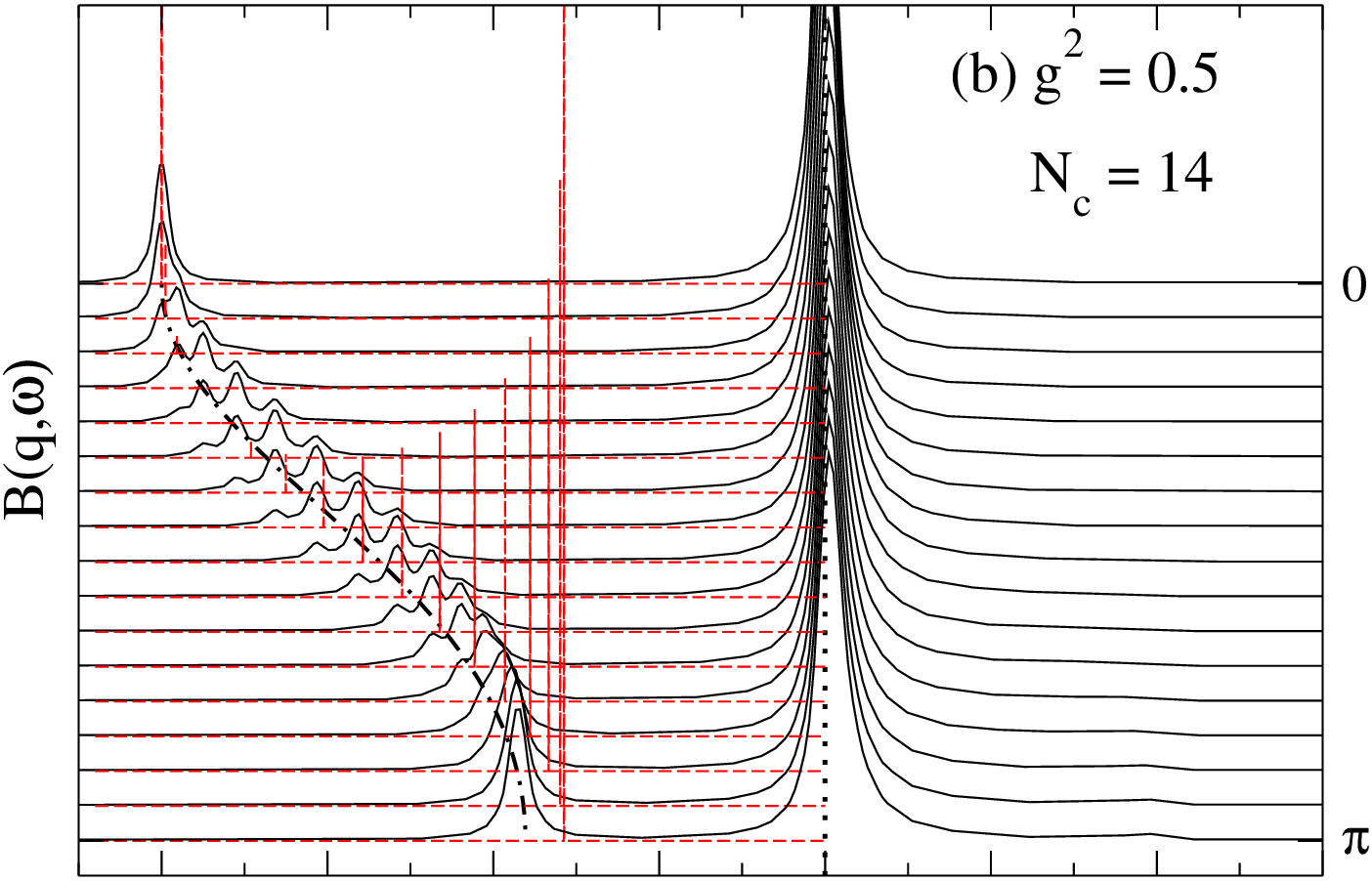}\\
  \includegraphics[width=0.52\textwidth]{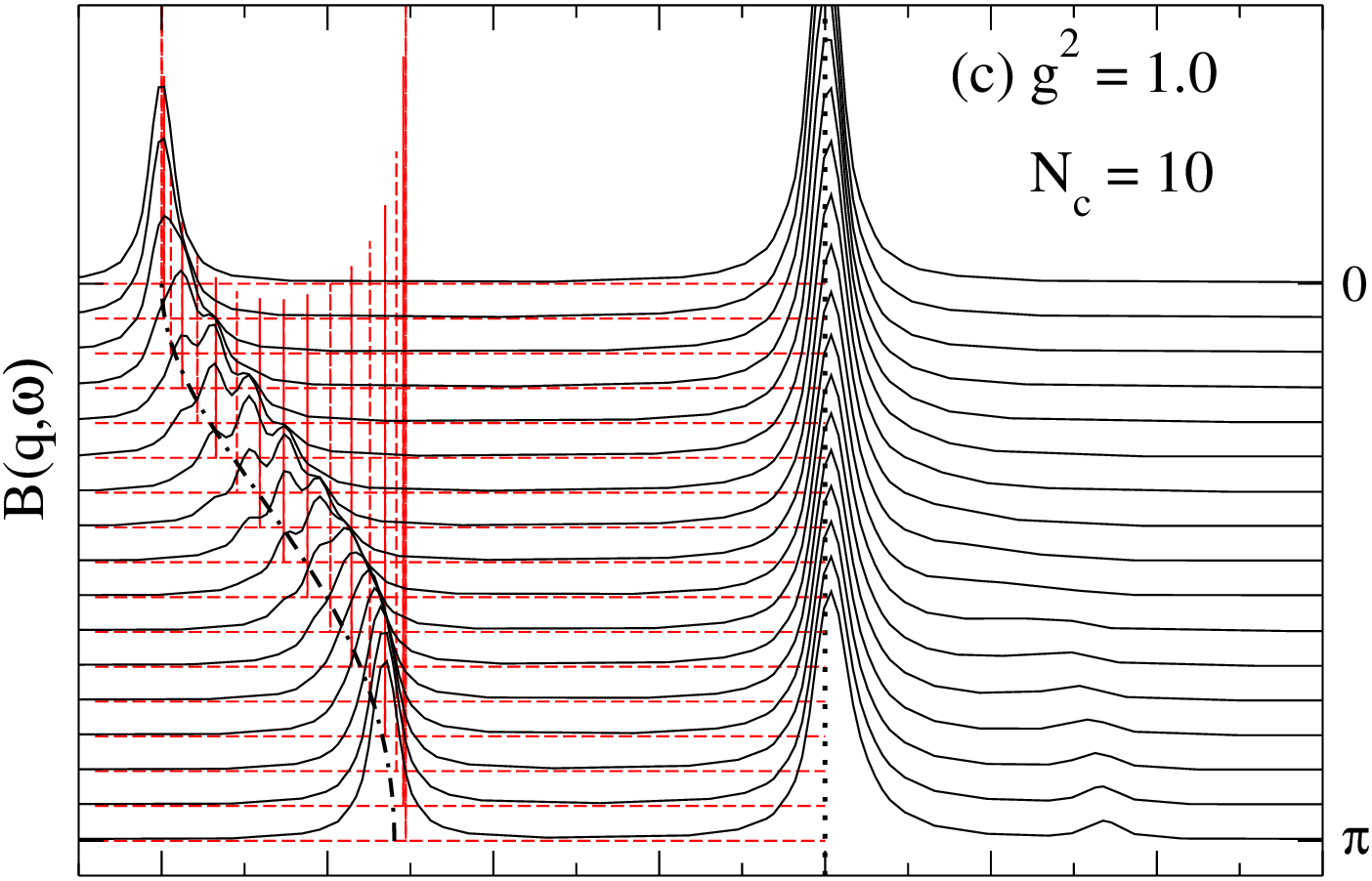}\\
  \includegraphics[width=0.52\textwidth]{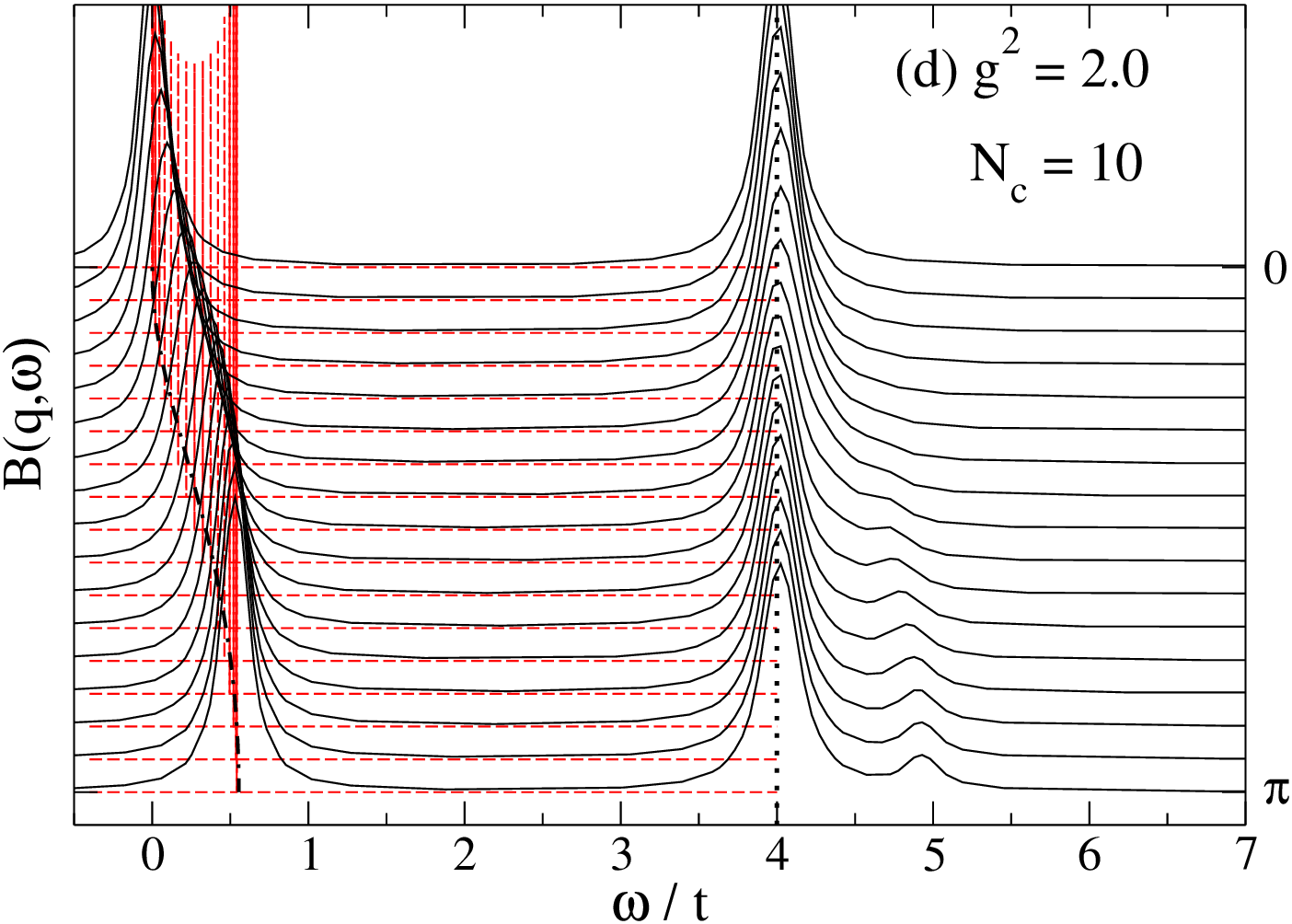}
  \caption{\label{fig:polaron_w4.0} (colour online) As in
    figure~\ref{fig:polaron_w0.4}, but for $\alpha=4$ and for different
    couplings $g^2$. WC (a) and SC (b)\,--\,(d) ($\dashed$, see text)
    analytical results are shown on a logarithmic scale.}
\end{figure*}

For a clearer representation, in the figures, we shall only show the non-trivial lower
excitations in the analytical results, \ie, $B(q,\om<\om_0)$ according to
equation~(\ref{eq:B<}) for WC, and only the second term on the rhs of
equation~(\ref{eq:Bqw_SC}) for SC, respectively. Furthermore, the analytical
data will be rescaled for better visibility (see captions of
figures~\ref{fig:polaron_w0.4} and~\ref{fig:polaron_w4.0}).

\paragraph{Weak coupling:}

If $g=0$, only the resonance excitation of phonons having the eigenfrequency
$\om_q=\om_0$ (the Holstein model~(\ref{eq:H}) neglects any dispersion of the
phonon branch) takes place, and the phonon spectral function is represented
by the delta function~(\ref{eq:B>}). A non-zero EP coupling connects the
lattice variables to the charge-carrier ones, giving rise to the
low-frequency (off-resonance) part of the spectral function, which reflects
the transitions to the excited polaronic states (of large or small polarons
for WC or SC, respectively). According to the WC analytical calculations of
section~\ref{sec:methods:wc}, this part of $B(q,\om)$ is given by
equations~(\ref{eq:im_pi}), (\ref{eq:B<}), and reflects the coherent part of
the electron spectral function lying in the frequency range $\om<\om_0$.  All this is confirmed in
figure~\ref{fig:polaron_w0.4}(a) for $\alpha=0.4$, \ie in the adiabatic
regime, which also shows the polaron band dispersion in the thermodynamic
limit from variational exact diagonalization \cite{BoTrBa99}.

Contrary to the adiabatic case shown in figure~\ref{fig:polaron_w0.4}(a), in
the WC anti-adiabatic case ($\alpha=4$) reported in
figure~\ref{fig:polaron_w4.0}(a), the lower excitation in $B(q,\om)$ remains
separated from the phonon line $\om=\om_0$, and corresponds to the entire
band of renormalized electron energies given---within our analytical
approach---in Sec.~\ref{sec:methods:wc} as the solution of
equations~(\ref{eq:band_WC})\,--\,(\ref{eq:wc_z_k}).

Both in the adiabatic (figure~\ref{fig:polaron_w0.4}(a)) and the
non-adiabatic (figure~\ref{fig:polaron_w4.0}(a)) WC cases, we find a very
good agreement of the WC approximation and the results from the cluster
approach and exact diagonalization, respectively, with only minor deviations
at large $q$ for $\alpha=4$. These deviations, also affecting the
$q$-dependence of the peak height in the analytical results, are a result of
the shortcomings of the method for $q$ lying near $0$ or $\pi$ (see
section~\ref{sec:analytical}).

An important point is that within the WC approximation, $B(q,\om)$ is
strongly suppressed for $q=0$ and $\om<\om_0$ due to the divergence of
equation~(\ref{eq:im_pi}). Hence, the peak in the numerical results is not
reproduced. In contrast, the SC approximation corresponds to undamped
quasiparticles (polarons) with a strong signal at $q=0$ (cf
equation~(\ref{eq:Bqw_SC})). Both these anomalies, connected with the
dispersion in the 1D electron (polaron) band, are a consequence of the
approximations used and have no physical relevance.

\paragraph{Intermediate coupling:}

The characteristic structure of $B(q,\om)$ consisting of the phonon line and
the low-energy part continues to hold even at stronger EP coupling.
Interestingly, for $\alpha=0.4$ at IC (figure~\ref{fig:polaron_w0.4}(b)), we
observe level repulsion between the weakly renormalized electron band and the
bare phonon excitation at some wave number $q_Y$---determined by the point in
$q$-space where the curves $\om=E_q-E_0$ and $\om=\om_0$ would intersect---as
in the case of the electronic spectrum \cite{HoAivdL03}.  For $\lambda=1$
(figure~\ref{fig:polaron_w0.4}(c)), the critical coupling for small-polaron
formation, the low-energy feature has already separated from the phonon line,
the latter being overlaid by an excited ``mirror band'' lying an energy
$\om_0$ above the polaron band.

The small-polaron cross-over for $\alpha=4$ is determined by the ratio $g^2$,
and occurs at $g^2=1$.  The phonon spectrum at this critical coupling is shown
in figure~\ref{fig:polaron_w4.0}(c).  We detect a clear signature of the
small-polaron band with renormalized half-width of about $0.70t$, in good
agreement with the SC result $\tilde{W}\approx0.74t$, but an order of
magnitude larger than in the adiabatic case of
figure~\ref{fig:polaron_w0.4}(c). In the latter, the SC approach predicts
$\tilde{W}\approx0.01t$, which is significantly smaller than the numerical
result of about $0.09t$. The fact that the analytical SC results are more
accurate in the non-adiabatic than in the adiabatic IC regime has been
pointed out before in \cite{LoHoFe06}.

Similar to figure~\ref{fig:polaron_w0.4}(c), figure~\ref{fig:polaron_w4.0}(c)
features a mirror image of the lowest polaron band---shifted by
$\omega_0$---with extremely small spectral weight, which is barely visible
for the number of Chebyshev moments---determining the energy
resolution---used here.

In the non-adiabatic case, small polarons exist even at intermediate EP
coupling (see \cite{HoNevdLWeLoFe04} and references therein). Therefore, as
underlined by the above comparison of the bandwidths, the SC approximation
agrees well with the exact results even for $g<g_\mathrm{c}$
(figure~\ref{fig:polaron_w4.0}(b)) and more than ever for $g=g_\mathrm{c}$
(figure~\ref{fig:polaron_w4.0}(c)).

By contrast, for $\alpha=0.4$, a large-polaron state exists for
$\lambda\approx\lambda_\mathrm{c}$, and both the WC (not shown) and the SC
approximation fail to reproduce the characteristic IC features of
figures~\ref{fig:polaron_w0.4}(b) and~\ref{fig:polaron_w0.4}(c).

\paragraph{Strong coupling:}

In the SC limit, the low-energy part is separated from the phonon line for
both the adiabatic and the anti-adiabatic case
(figures~\ref{fig:polaron_w0.4}(d) and~\ref{fig:polaron_w4.0}(d)), and the
full spectral function has the form given by equation~(\ref{eq:Bqw_SC}).
Moreover, the effect of the polaron band-narrowing is more pronounced for the
adiabatic case, as the small-polaron band at fixed $\lambda$ has the
half-width $\tilde{W}=W\rme^{-2\lambda/\om_0}$. In fact, there is no visible
dispersion in the lower or upper band in figure~\ref{fig:polaron_w0.4}(d).
As expected, for these parameters, the SC approximation fits well the
numerical results.

Another general feature related to the $\om_0$-dependence of the spectral
function becomes apparent if we compare the heights of peaks in
figures~\ref{fig:polaron_w0.4} and~\ref{fig:polaron_w4.0} (the ordinate scale
in figure~\ref{fig:polaron_w0.4} is about a factor of five larger than that
in figure~\ref{fig:polaron_w4.0}). This dependence reflects the fact that the
transition probability at low temperatures is proportional to the density of
polaron states in the coherent band. The latter quantity increases with
decreasing bandwidth, and the pronounced difference between adiabatic and
anti-adiabatic spectral functions in the SC limit is evident from
equation~(\ref{eq:Bqw_SC}), where the height of the peaks is dominated by the
factor $1/\tilde{W}\propto\rme^{2\lambda/\om_0}$.

We conclude that the WC approximation based on the second-order electron
spectral function, and the Hartree approximation for the SC small-polaron
limit are able to grasp the main qualitative features of the phonon spectral
function across the range of model parameters. However, as discussed in
section~\ref{sec:intro}, the numerical calculations do not represent exactly
the limit of negligible charge-carrier concentration, owing to the restricted
cluster volume. A direct comparison with results obtained by carrying out the
limiting procedure for the analytical formulas in
section~\ref{sec:analytical} would be possible only for $N\to\infty$.

On the other hand, no restrictions
concerning the charge carrier concentration were imposed in deducing
equation~(\ref{eq:pi3}) in section~\ref{sec:methods:wc} and
equation~(\ref{eq:D_SC}) in section~\ref{sec:methods:sc}.  Consequently, the
analysis of the latter equations outlined in section~\ref{sec:analytical}
permits us to discuss the additional features of spectra for non-negligible
carrier concentrations revealed by the numerical results. First, the
discussion in section~\ref{sec:methods:wc} strongly suggests that the
low-energy peaks of finite width correspond to the solutions of
equations~(\ref{eq:im_Pi})\,--\,(\ref{eq:om}) at fixed wavevector $q$ if
$\mu$ lies above the bottom of the electron (polaron) band. Second, at finite
concentrations, the contributions of the incoherent part $\Aic$ of the
electron (polaron) spectral function to $\Im\Pi(q,\om)$ are not negligible
and, in this way, additional maxima above $\om=\om_0$ occurring in the
numerical results may be understood as originating from phonon-assisted
processes implied in $\Aic$.  Finally, the non-zero $\Re\Pi(q,\om)$ in
general causes a shift of the bare phonon line away from $\om=\om_0$ but,
according to numerical results, the latter is not very pronounced in the WC
and SC cases.

%
%
%%%%%%%%%%%%%%%%%%%%%%%%%%%%%%%%%%%%%%%%%%%%%%%%%%%%%%%%%%%%%%%%%%%%%
\section{Summary}\label{sec:conclusions}
%%%%%%%%%%%%%%%%%%%%%%%%%%%%%%%%%%%%%%%%%%%%%%%%%%%%%%%%%%%%%%%%%%%%%
%
%
%

We have presented results for the phonon spectral function of the Holstein
polaron in all relevant parameter regimes obtained by a reliable and
systematic cluster approach similar to cluster perturbation theory. The
characteristic features of the spectra have been discussed and successfully
related to analytical self-energy calculations valid at weak and strong
coupling, respectively. As far as a direct comparison is possible,
our findings are in agreement with previous work on the phonon dynamics.

In particular, we have pointed out the important
differences between weak, intermediate and strong coupling, on the one hand,
and between the adiabatic and the anti-adiabatic regime, on the other hand.
As revealed by the analytical results, the phonon spectra of the Holstein
polaron are dominated by the bare, unrenormalized phonon line and the
renormalized polaron band dispersion. At intermediate coupling, additional
features such as level repulsion and mirror images of the polaron band have
been observed.  Together with previous studies of the electron spectral
function and the renormalization of phonon energies, this work
provides a fairly complete picture of the spectral
properties of the one-dimensional Holstein polaron, which has been in the
focus of intensive investigations over several decades due to the wide-spread
relevance of polaron physics.

%%%%%%%%%%%%%%%%%%%%%%%%%%%%%%%%%%%%%%%%%%%%%%%%%%%%%%%%%%%%%%%%%%%%%
\ack

This work was supported by HPC-Europa, the Deutsche Forschungsgemeinschaft
through SPP1073, the DFG and the Academy of Sciences of the Czech Republic
(ASCR) under Grant Nr.~436 TSE 113/33/0-2. We would like to thank G Wellein
for valuable discussion.

%%%%%%%%%%%%%%%%%%%%%%%%%%%%%%%%%%%%%%%%%%%%%%%%%%%%%%%%%%%%%%%%%%%%%

%%%%%%%%%%%%%%%%%%%%%%%%%%%%%%%%%%%%%%%%%%%%%%%%%%%%%%%%%%%%%%%%%%%%%%%%%%%%%
%%%%%%%%%%%%%%%%%%%%%       BIBLIOGRAPHY              %%%%%%%%%%%%%%%%%%%%%%%
%%%%%%%%%%%%%%%%%%%%%%%%%%%%%%%%%%%%%%%%%%%%%%%%%%%%%%%%%%%%%%%%%%%%%%%%%%%%%

\section*{References}

%\bibliography{../bibliography/bibliography}
%\bibliographystyle{prsty}

\end{document}